\documentclass[5p,twocolumn,number,sort&compress]{elsarticle}

\usepackage{graphicx}

\begin{document}

\begin{frontmatter}

\title{Studying Pulsed Laser Deposition conditions for Ni/C-based 
multi-layers}
% author list
\author{Tjeerd R.J. Bollmann}
\address{University of Twente, Inorganic Materials Science, MESA$^+$ 
Institute for Nanotechnology, P.O. Box 217, NL-7500AE Enschede, The Netherlands}
% \ead{t.r.j.bollmann@utwente.nl}

\date{\today}

\begin{abstract}
% A\\B\\S\\T\\R\\A\\C\\T\\
Nickel carbon based multi-layers are a viable route towards future hard X-ray 
and soft $\gamma$-ray focusing telescopes. 
Here, we study the Pulsed Laser Deposition growth conditions of 
such bilayers by Reflective High Energy Electron Diffraction, X-ray 
Reflectivity and Diffraction, Atomic Force Microscopy, X-ray Photoelectron 
Spectroscopy and cross-sectional Transmission Electron Microscopy analysis, 
with emphasis on optimization of process pressure and substrate temperature 
during growth.
The thin multi-layers are grown on a 
treated SiO substrate resulting in Ni and C layers with surface roughnesses 
(RMS) of $\leq$0.2~nm.
Small droplets resulting during melting of the targets 
surface increase the roughness, however, and can not be avoided.
The sequential process at temperatures beyond 300$^\circ$C results 
into intermixing between the two layers, being destructive for the reflectivity 
of the multi-layer.
\end{abstract}

\end{frontmatter}
% \maketitle

%% Introduction 
\section{Introduction}
\begin{figure}
	
\includegraphics[width=8cm]{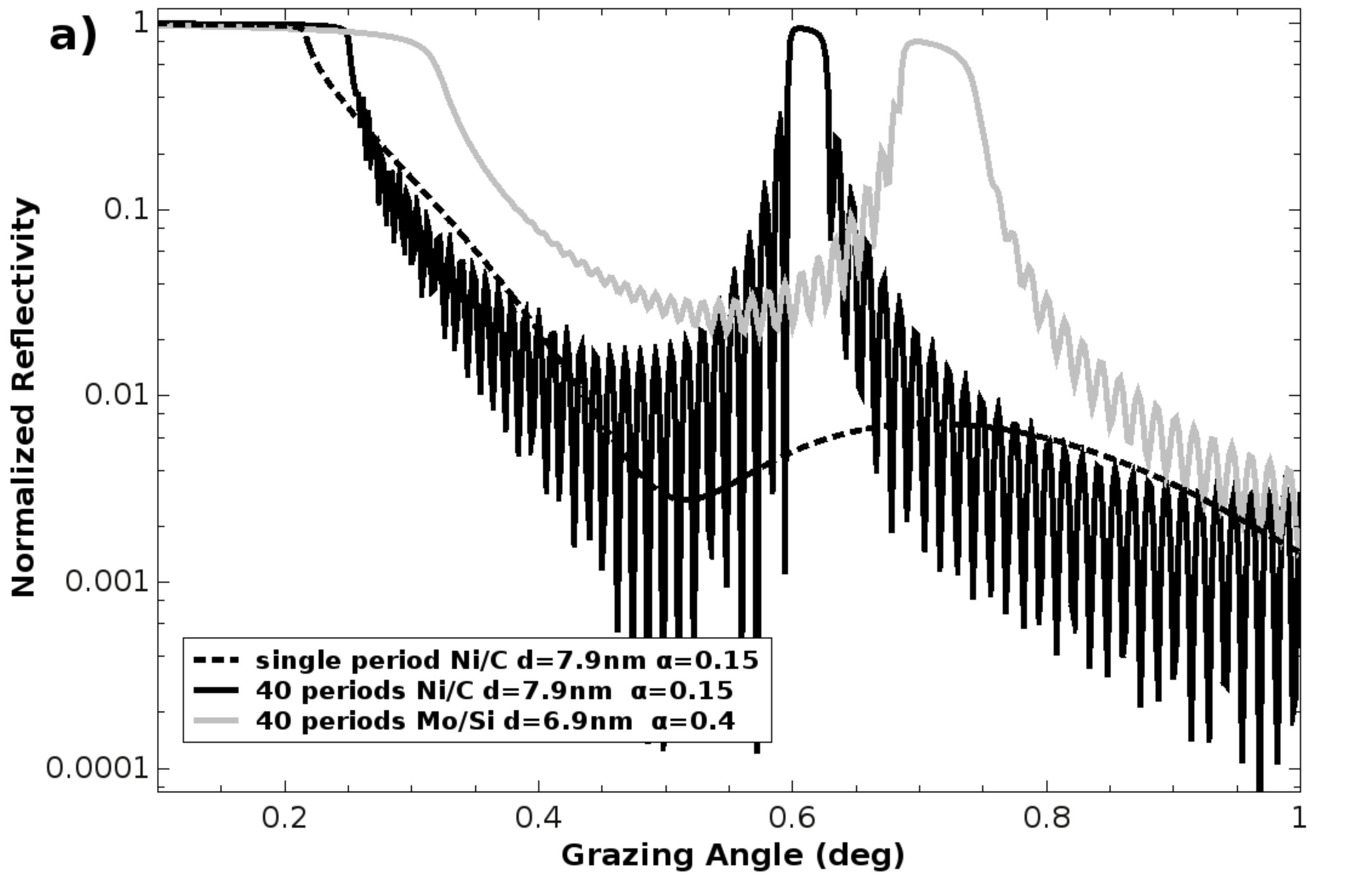}
\includegraphics[width=8cm]{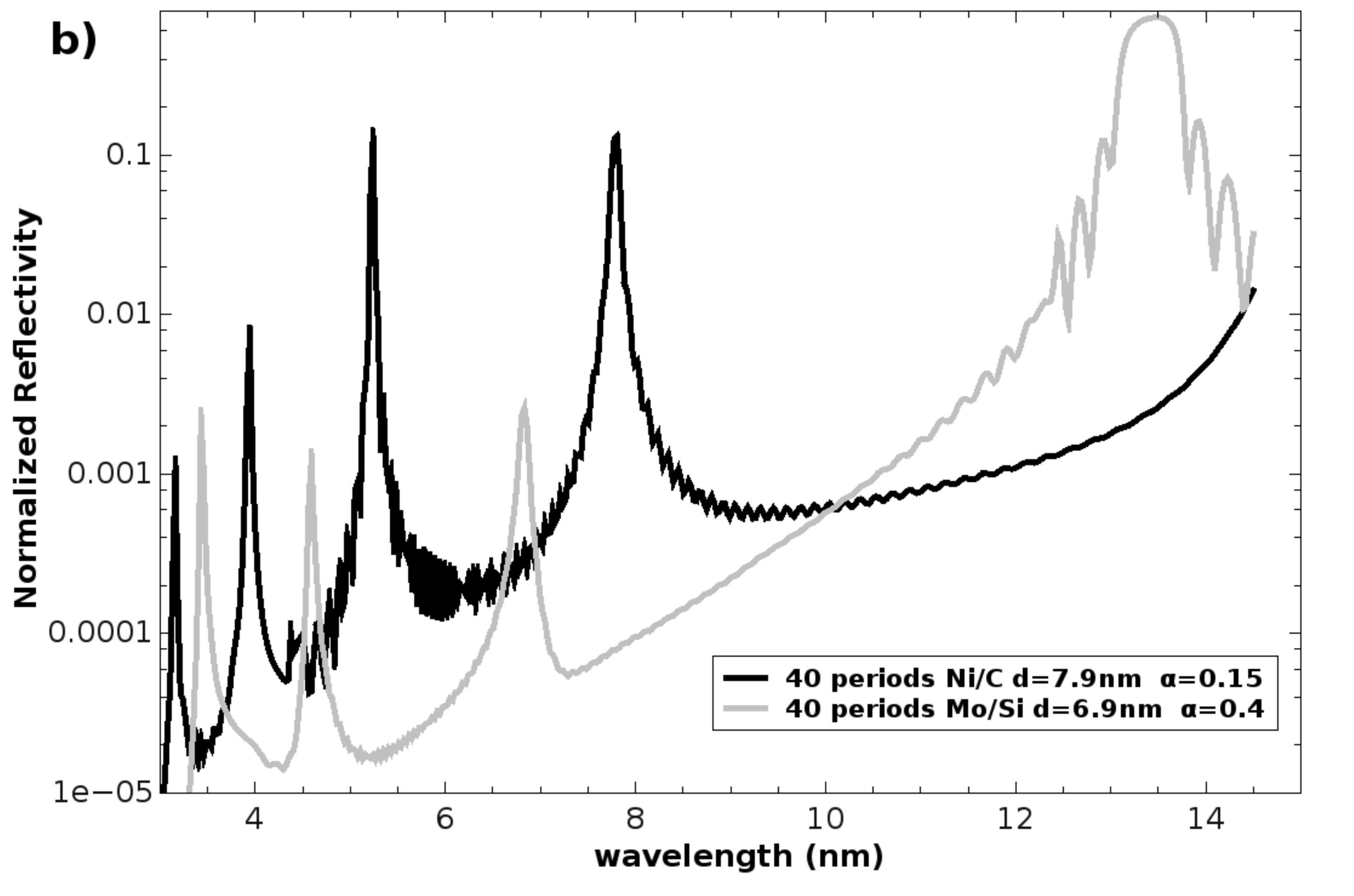}
\caption{
(a) Calculated X-ray reflectivity curves at 8.05~keV (Cu K-$\alpha$) of a Ni/C 
multi-layer with 1 and 40 pair(s) of 1.2~nm Ni and 6.7~nm C layers, and of 
a Mo/Si 
multi-layer with 40 pairs of 2.76~nm Mo and 4.14~nm Si layers.
(b) Calculated X-ray curves of the Ni/C multi-layer and Mo/Si multi-layer 
both with 40 pairs at wavelengths of 3 to 14.5~nm.}
\label{fig:0}
\end{figure}

Bragg mirrors, consisting of multiple thin layers of dielectric material, can 
by carefull design reflect for specific wavelenghts of electromagnetic 
radiation.
Where single layer mirrors are rather limited in that the incoming grazing 
angles should be below or close to the critical angle of that material, 
multi-layer mirrors allow for larger angle windows.
Multilayered thin film mirrors efficiently reflecting X-rays, nowadays find 
their application towards a variety of applications such as EUV 
photolitography, X-ray microscopy, synchrotron radiation, plasma physics and 
astrophysics. For EUV photolitography working around a 13~nm wavelength, 
typically Mo/Si multi-layers having a reflectivity of about 60\% find 
good use. At lower wavelenghts, however, their reflectivity becomes 
very low, a region where Ni/C multi-layers have higher reflectivity. 
Spiga et al. \citep{Spiga2004}
e.g. measured a reflectivity of 95\%.
For current hard X-ray focusing telescopes, the use of Pt and the corresponding 
K adsorption edge leads to an upper energy limit of $\approx$79.4~keV. The 
usage 
of a different heavy element in the multi-layers, such as Ni, can result in an 
extended energy bandwidth into the soft $\gamma$-ray range as Ni does not have 
adsorption edges in the hard X-ray/soft $\gamma$-ray range.
Especially depth-graded Ni/C multi-layers find their use in focusing optics in 
astronomical telescopes that aim for hard X-ray and $\gamma$-ray observatories 
with typical energies of 158~keV and 511~keV, important to study respectively 
the Ni decay and positron-electron annihilation in supernovae \citep{Buis2006}.
In Fig.~\ref{fig:0} we plotted the calculated X-ray reflectivity curves at 
8.05~keV versus 
grazing angle and versus wavelength using Fresnel equations and Henke's optical 
data \citep{Henke1988}. Calculations are shown for a single Ni/C multi-layer 
and 
a 40 
pairs Ni/C multi-layer, both with a Ni film thickness of 1.2~nm and a C film 
thickness of 6.7~nm, as suggested for an optic design for use in hard X-ray and 
soft $\gamma$-ray astronomy applications using silicon pore optics 
\citep{Collon2016}. 
% The high reflectivity of 19 pairs of multi-layers at a grazing angle of 0.6 
% degrees, is experimentally verified in Fig. 5 of Ref.~\cite{Spiga2004}.
%
For comparison 
we plotted the calculated X-ray reflectivity curves for a 40 pairs Mo/Si 
multi-layer of 2.76~nm Mo and 4.14~nm Si. Notice the superior 
reflectivity for Ni/C multi-layers at grazing angles in comparison to the Mo/Si 
multi-layers, see Fig.~\ref{fig:0}(a), and the vanishing reflectivity at lower 
wavelenghts of the Mo/Si multi-layers in comparison to the Ni/C multi-layer, 
see 
Fig.~\ref{fig:0}(b).

Designing towards shorter periodic wavelenghts, however, reduces the 
experimentally 
observed reflectivity significantly in comparison to the anticipated 
theoretical reflectivity. This is anticipated to result from the substantial 
interface roughness, caused by crystallization of Ni and is found to occur 
below a thickness of 2~nm followed by interdiffusion between adjacent Ni and C 
layers worsening the observed interlayer roughness 
\citep{Krawietz1995, Takenaka1996, Dietsch1998, Georgescu1999, Peng2016}. In 
order to 
reduce the roughness between adjacent layers, Pulsed Laser Depositing (PLD) 
these films is found to result in smoother films as the deposited particles 
having higher kinetic energy of several 100~eV can smoothen the Ni/C interface. 

Here, we investigate the feasibility of producing such thin 
multi-layers by the use of Pulsed Laser Deposition on a silicon-oxide (SiO) 
substrate containing a native oxide. Pulsed Laser 
Deposition is a well studied technique to aid in the synthesis of 
multicomponent and complex thin films. In comparison to conventional vapor 
deposition techniques, PLD provides additional parameters thereby in principle 
giving more flexibility to the (control of the) growth process of thin films. 
Besides this, PLD is known for its high supersaturation in growth, resulting in 
a high nucleation density resulting in (ultra)smooth thin film growth 
\citep{Blank2014}, a necessity for the application described before. Although a 
lot of work is done on the PLD growth of (complex) oxides \citep{Blank2014}, 
only 
little studies are done on the initial stages of the growth of metallic thin 
films \citep{Krebs1993, Krebs1995, Lunney1995, Faehler1997, Krebs1997, 
Doeswijk2004}. In this work, we study the evolution of metallic thin films 
grown 
by PLD with emphasis on the process variables. 
% Studying the parameters of 
% influence, we aim at controlling the metallic thin film quality as it could 
% serve as a next generation coating with extremely low roughness for application 
% in future hard X-ray and soft $\gamma$-ray optics.
% 
By studying the growth at varying conditions, we seek the optimal PLD growth 
conditions for Ni/C thin film multi-layers as well as studying the influence on 
the resulting reflectivity, as these multi-layers could serve as a next 
generation coating for application in future hard X-ray and soft $\gamma$-ray 
optics.

In this paper, we therefore start by the growth and characterization of 
individual Ni and C films in order to optimize the individual growth conditions. 
We then stack these films at elevated temperature to 
test their thermal stability. As analysis reveals the intermixing of Ni and C 
at elevated temperatures, we summarize by analysis of pairs of Ni/C 
multi-layers grown at RT conditions, discussing the experimentally 
observed reflectivity in view of the theoretically predicted one. This brings us 
to conclusions and outlook towards future applications.

%% Experimental
\section{Experimental details}
For comparison to commercially available silicon-oxide pore optics 
\citep{Collon2016}, we diced a native silicon-oxide wafer in samples of 
5$\times$5~mm$^2$ which were then ultrasonically degreased in 
acetone and ethanol prior to insertion into the vacuum setup. The SiO 
sample was glued by 
silver-paste to a sample-holder that could be heated in-situ by laser heating. 
The sample temperature was measured by a pyrometer at the backside of the 
sample-holder, estimated to result in an error of $\pm$50$^{\circ}$C.

A PLD system from Twente Solid State Technology (TSST) {B.V.} was used in 
combination with 
a 248~nm excimer laser
with a pulse duration of 25 ns to ablate material from the targets: 
single crystal Ni(111) and Highly Oriented Pyrolytic Graphite (HOPG) of ZYA 
grade. A rectangular mask was 
used to create a well-defined and homogeneous laser profile on the target.
The target was positioned 50~mm from the substrate. The laser fluence was 
controlled using a variable beam attenuator. By applying a magnification of 
11.5, we achieved a laser energy density of 6~J/cm$^2$ for the ablation 
of C and 7.5~J/cm$^2$ for the ablation of Ni, using a spot-size of 
0.6~mm$^2$. Both were scanned during deposition over a width of 5~mm at a speed 
of 0.5~mm/second. The background gas pressure was in 
the range of 10$^{-8}$ mbar. 

The growth rate per pulse was obtained by determination of the Kiessig 
fringes in the resulting X-ray reflectivity curve for a thick film grown. For 
Ni 
we determined a growth rate of 1.9$\times$10$^{-3}$~nm/pulse where growth was 
performed at a repetition rate of 8~Hz. The growth rate for C was found to be 
1.8$\times$10$^{-2}$~nm/pulse at a repetition rate of 10~Hz. 

During deposition the surface structure was 
monitored using Reflective High Energy Electron Diffraction (RHEED) operated at 
30~keV, at pressures up to 0.3~mbar, enabled 
due to differential pumping. Subsequently, X-ray reflectivity (XRR) and 
diffraction (XRD) at a Cu K-$\alpha$ wavelength 
was used for reflectivity and structural characterization of the multi-layers. 
By use of ex-situ tapping mode atomic force microscopy (TM-AFM) the resulting 
surface roughness ($\sigma_{RMS}$) was determined. The surface roughness (RMS) 
of the 
pristine SiO samples was found to be $\approx$0.58~nm.

\section{Results and discussion}
To study and optimize the PLD growth conditions for individual Ni and C films, 
typical parameters in the PLD method should be considered; the laser fluence, 
giving the target 
particles their initial kinetic energy; the laser repetition rate, dictating 
the growth rate; the process pressure, reducing the kinetic energy of the 
particles arriving on the substrate as well as the plasma shape 
\citep{Groenen2015}; and the substrate temperature, enhancing diffusivity 
during 
the growth. Here, we focus on the role of process pressure and substrate 
temperature upon growth, as they mainly govern the roughness of the resulting 
film as for 
its reflective properties of the multi-layer we are not interested in obtaining 
highly crystalline films.

As is described in literature for the growth of thin graphite films, an 
increase 
in laser energy density results in higher quality films \citep{SarathKumar2013, 
SarathKumar2013b}. Besides this, a thermal treatment is known to increase the 
thin film quality \citep{SarathKumar2013b, Tite2014}. As a low surface 
roughness 
is of utmost importance for future applications, we deposited thin graphite 
films at the described conditions, an ablation threshold of 6~J/cm$^2$, for 
increasing process pressures of respectively 0.027, 0.05 and 0.1~mbar of Ar, as 
an increased process pressure typically leads to more 2D growth for PLD growth 
\citep{Sturm2000,Scharf2002,Scharf2004,Groenen2015}. For PLD films grown at 
process pressures beyond 0.027~mbar, we find however a sharp increase in 
roughness.
Therefore, the process 
pressure throughout this paper was 0.027~mbar of Ar by influx of a 40 
standard cubic centimeter (SCCM) Ar flow and subsequent control on the pump 
rate by positioning the gate valve between pump and vacuum 
chamber. 

In comparison to the PLD of oxides, the ablation of metal targets requires 
ablation thresholds typically an order of magnitude higher 
\citep{Doeswijk2004}. 
However, too large ablation thresholds in combination with a too low process 
pressure results in droplet formation at growth \citep{Krebs1993, Krebs1995, 
Lunney1995, Krebs1997, Faehler1997}. A study on the microstructural evolution 
of Ni films upon increasing ablation threshold \citep{Kumar2008} revealed an 
increased clustering and denser packing of nanoparticles. In order to obtain 
sufficiently high deposition rates in high vacuum, laser fluences of more than 
5~J/cm$^2$ are a necessity. Besides this, it should be noted that in 
order to reduce the droplet density, a sweet spot of ~7-9~J/cm$^2$ 
exists, as discussed in Fig.~5 of Ref. \citep{Faehler1997}. Making use of an 
ablation threshold of 7.5~J/cm$^2$ and a 
process pressure of 0.042~mbar to reduce the kinetics of the arriving 
particles, 
we obtain low roughness Ni surfaces which show no preferential crystallographic 
orientation. 

\begin{figure}
\includegraphics[width=8cm]{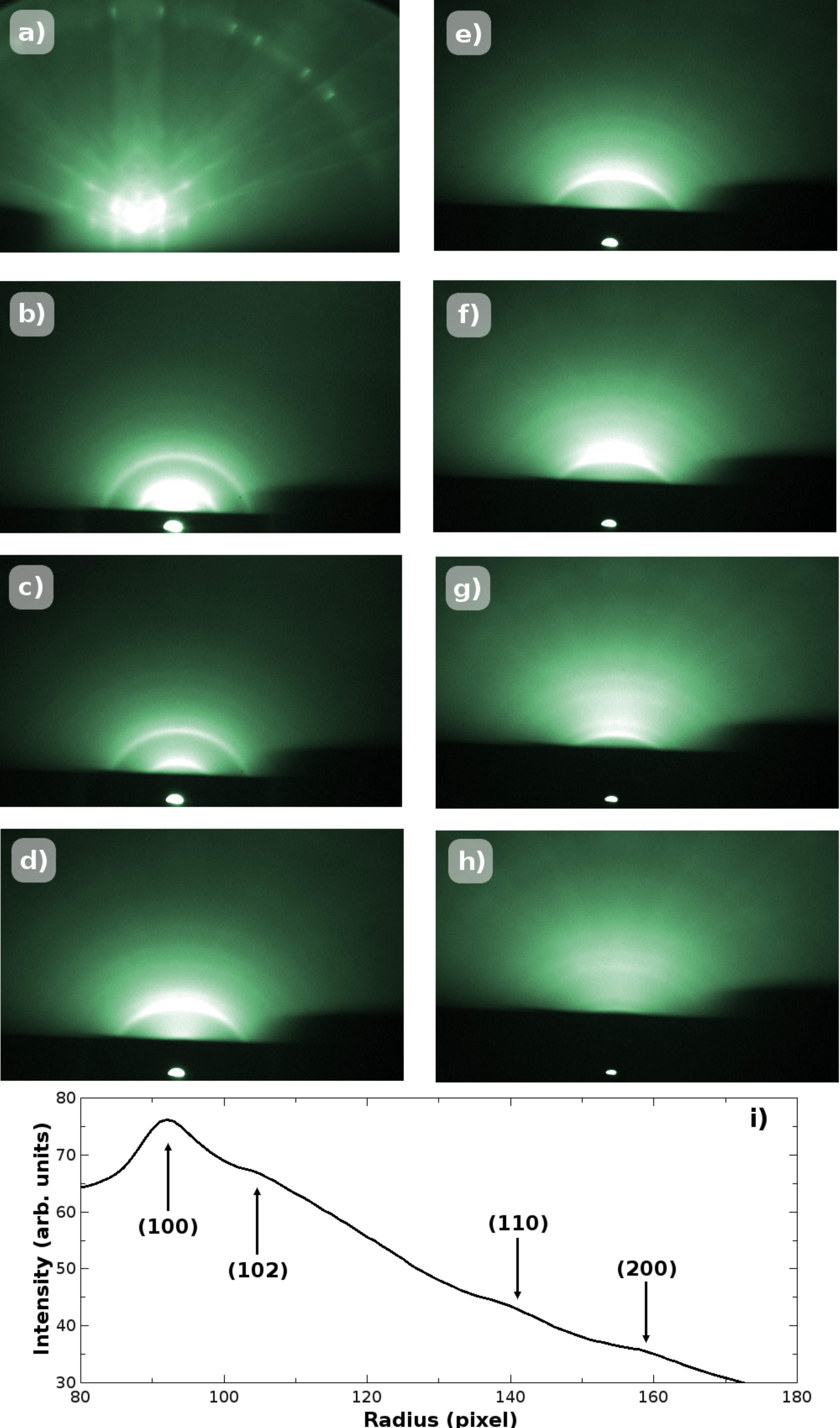}
\caption{(color online) (a) RHEED pattern of the SiO substrate. (b) 
Transmission RHEED pattern of as-deposited C film at an incident angle of 
1$^\circ$ (b), 1.5$^\circ$ (c), 2$^\circ$ (d), 2.5$^\circ$ (e), 3$^\circ$ (f), 
3.5$^\circ$ (g), 4$^\circ$ (h). All RHEED patterns where taken at a primary 
electron energy of 30~keV. (i) The integral of the radial intensity 
distribution of the as grown C film. The labeled arrows 
indicate the expected positions of the Debye-Scherrer rings 
calculated from powder diffraction intensities.}
\label{fig:3}
\end{figure}

In order to study the quality of the mirror, which depends on the quality of 
every single layer deposited, we started by analyzing single layers of Ni and C 
deposited by PLD.
Before deposition, the RHEED pattern reveals a clean SiO substrate pattern as 
shown in Fig.~\ref{fig:3}(a). Upon deposition of C at 
RT, a continuous decay in the intensity of the substrate spots appears. For 
thicker films, clear transmission RHEED patterns of multiwalled carbon 
nanotubes appear, see e.g. Fig.~\ref{fig:3}(b). The 
transmission pattern was checked by tilting the sample as well as varying the 
angle of incidence with respect to the surface, see Fig.~\ref{fig:3}(c-h). By 
this, substrate features should reveal position changes as they are very 
sensitive to tilt, in contrast to transmission features which keep their 
position.

In order to calculate the preferred interplanar distances within 
the C film, the preferred Debye-Scherrer ring radius is calculated 
by integrating the radial intensity distribution along the periphery of circles 
around the RHEED central spots for increasing radius.
% , see also Supp. Mat. 
Fig.~\ref{fig:3}(i) shows the integral of the radial intensity distribution of 
Fig.~\ref{fig:3}(c). 
Similar to RHEED images described for randomly oriented multiwalled carbon 
nanotube samples, the radial profile of the diffraction intensity is in 
correspondence with graphites lattice parameters $a$ and $2b$ as described in 
detail in Ref.~\citep{Drotar2001}.

%Figure 3: RHEED image analysis, ring pattern, no preferential 140821 met matlab

For Ni films grown, we only observe a continuous decay in the RHEED intensity 
of the 
substrate spots, however, no formation of a transmission RHEED pattern. This is 
also the case for thicker Ni films, indicative of the amorphous 
microstructure of the layer, a necessity for efficient mirrors.

\begin{figure}
\includegraphics[width=8cm]{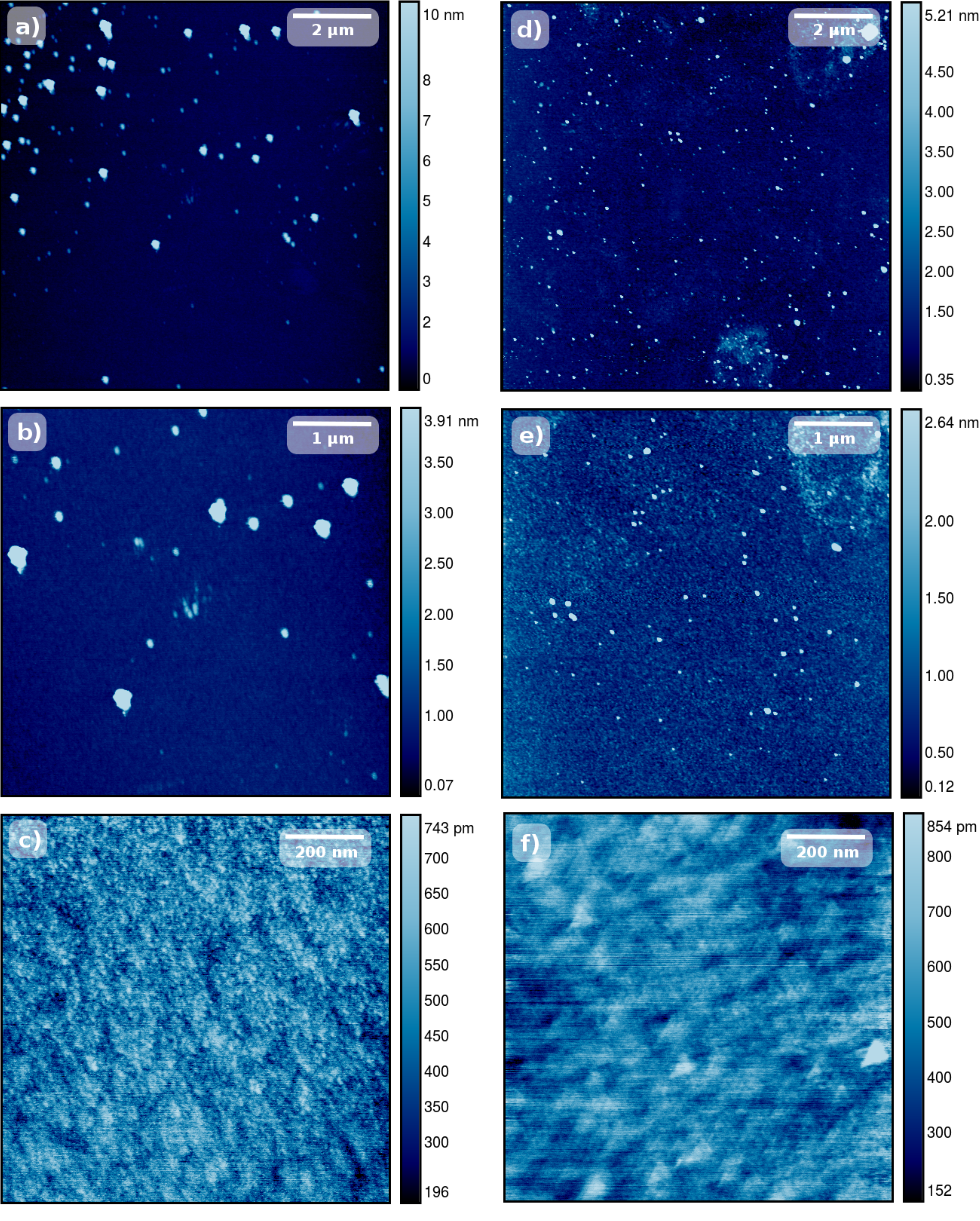}
\caption{(color online) Topographic TM-AFM images of a 1.2~nm deposited Ni film 
on the SiO substrate of 10$\times$10$\mu$m$^2$ (a), 5$\times$5$\mu$m$^2$ (b) 
and 1$\times$1$\mu$m$^2$ (c), and of a 6.7~nm thick deposited C film on the SiO 
substrate of 10$\times$10$\mu$m$^2$ (d), 5$\times$5$\mu$m$^2$ (e) and  
1$\times$1$\mu$m$^2$ (f). }
\label{fig:1}
\end{figure}

In Fig.~\ref{fig:1}(a-c), we show topographic TM-AFM images of the grown 1.2~nm 
thick Ni and 6.7~nm thick C films. The different AFM 
image sizes reveal roughnesses (RMS) $<$0.1-1.4~nm. In comparison to 
magnetron-sputtered Ni films \citep{Peng2016}, where the roughness increases 
for 
decreasing film thickness, the roughness at the interface seems rather ill 
defined. We should however take into account that during the PLD processing of 
metals, the formation of spherical droplets with sizes in the range of 
0.1-0.2~$\mu$m are often reported \citep{Faehler1997, Largeanu2011}.
From a grain analysis on the AFM images shown in Fig.~\ref{fig:1} we find 
spherical droplets having similar dimensions. 
Although larger sized 
droplets can easily be avoided, these 
smaller droplets result from the fast heating and cooling occuring 
during melting of the targets surface and can not be completely avoided. 
Literature suggests the use of some kind of velocity filter 
in order to obtain droplet free films \citep{Faehler1997}. By marking these 
small 
droplets and excluding them from the analysis, we find roughnesses 
(RMS) in the range of $<$0.1-0.2~nm. If one would be able to avoid the droplets 
resulting from the heating and cooling at the target, one would obtain PLD 
films superior to magnetron-sputtered films.

For the roughness (RMS) of the C films, see Fig.~\ref{fig:1}(d-e), we find 
values between 0.2-0.8~nm. For these films, we observe an increased number of 
small droplets in line with the observations for the Ni films. Excluding the 
droplets from the image analysis, reveals a roughness (RMS) of 0.2~nm, in 
agreement with the observations for the thickness dependent roughness (RMS) of 
magnetron-sputtered C films \citep{Peng2016}. 

%Figure 1: 	AFM RMS of a 1.2nm thick Ni film column 1
%		AFM RMS of a 6.7nm thick C film column 2
% discuss droplet formation, and C is in line with Scientific reports, but Ni 
% does much better

\begin{center}
\begin{table}
\begin{tabular}{l|l|l}
AFM image size & $\sigma_{RMS}$ Ni film & $\sigma_{RMS}$ C film\\
\hline
10$\times$10$\mu$m$^2$ & 0.23nm (1.36nm) & 0.24nm (0.78nm)\\
\hline
5$\times$5$\mu$m$^2$ & 0.14nm (0.91nm) & 0.22nm (0.51nm)\\
\hline
1$\times$1$\mu$m$^2$ & 0.09nm & 0.11nm
\end{tabular}
\label{tab:1}
\caption{Roughness (RMS) measured for different AFM image sizes, excluding 
masked nm-sized particles. Between brackets, the roughness (RMS) including the 
entire image.}
\end{table}
\end{center}

%% combining both
Having found PLD conditions to grow Ni and C films, both, with surface 
roughnesses $\leq$0.2~nm for the required thicknesses of respectively 1.2 and 
6.7~nm, we stack both films, thereby changing the growth conditions in 
sequence. For the Ni film we use an ablation threshold of 7.5~J/cm$^2$ and a 
substrate temperature at RT, where for the following C film we use an ablation 
threshold of 6~J/cm$^2$ and vary the growth temperature 
as a thermal treatment is known to increase the thin film quality 
\citep{SarathKumar2013b, Tite2014} and more importantly, to test the thermal 
stability of the multi-layer as it is required in astronomy applications 
\cite{Collon2016}.

\begin{figure}
\includegraphics[width=8cm]{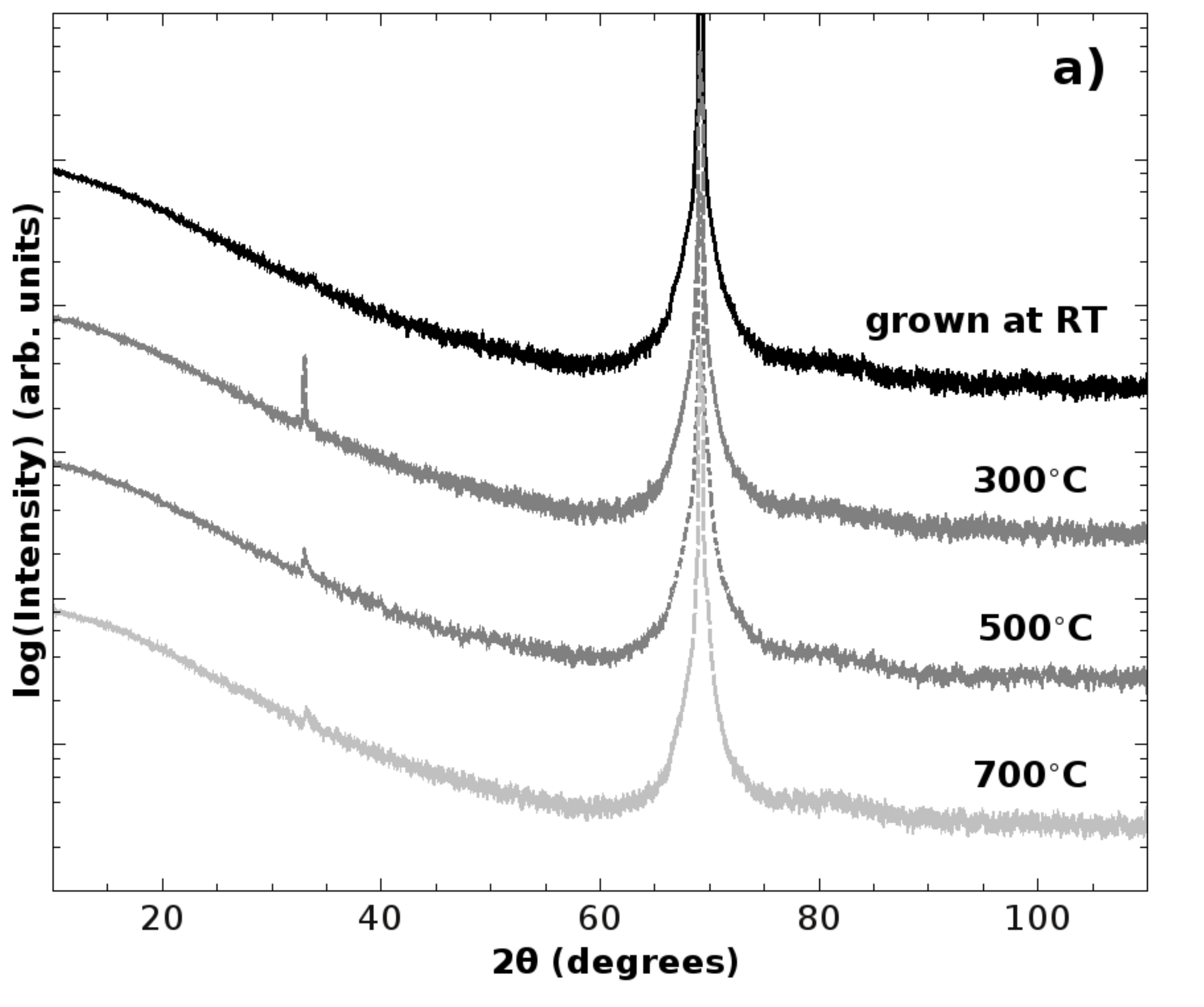}
\includegraphics[width=8cm]{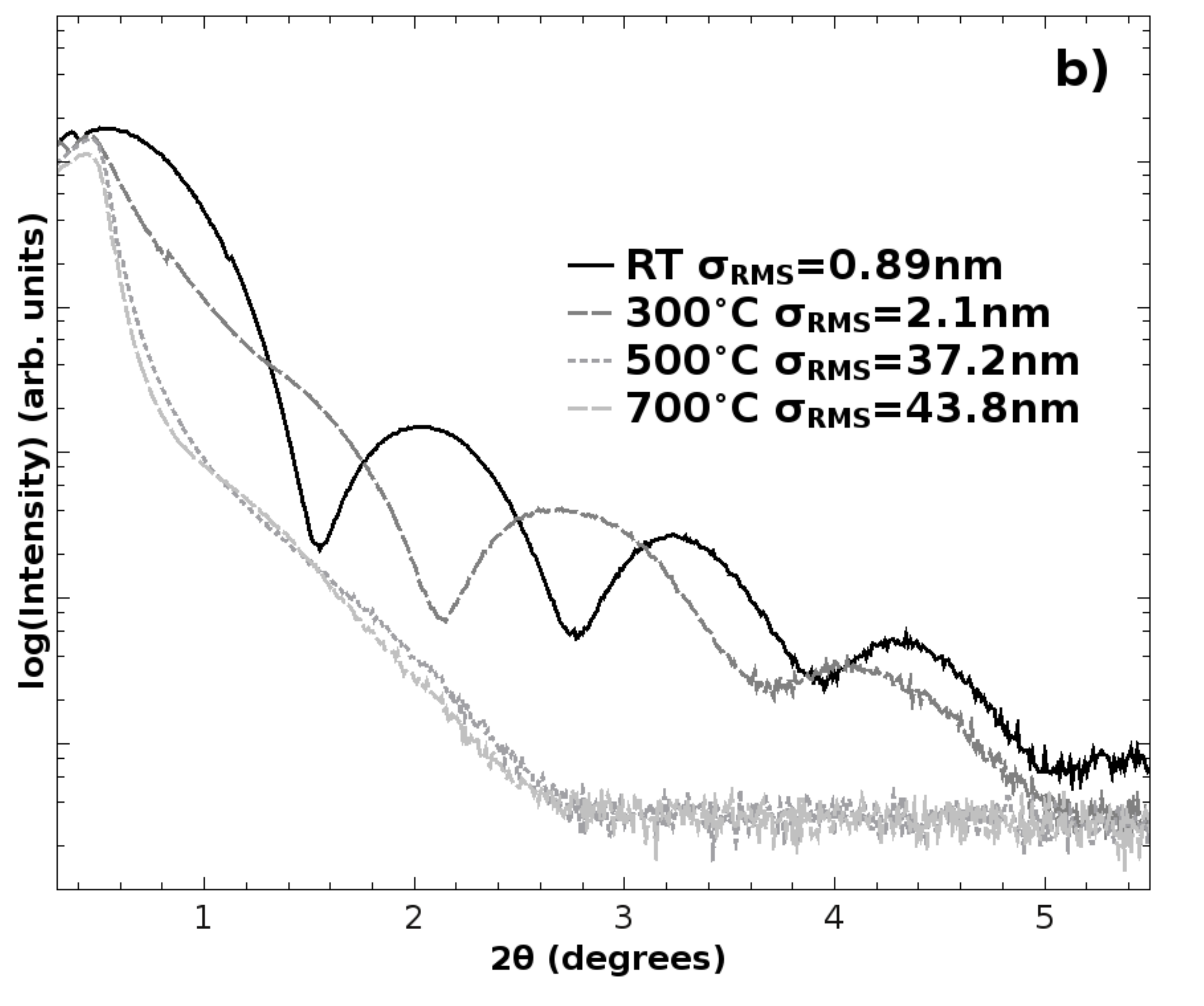}
\caption{(a) X-ray diffraction curves of a Ni/C multi-layer, 
20~nm Ni and 75~nm C, grown at increasing temperature. Spectra are vertically 
offset for presentation. (b) The 
corresponding X-ray reflectivity curves. Beyond a growth 
temperature of 300$^{\circ}$C, the Kiessig fringes are not present.}
\label{fig:2}
\end{figure}

To test the thermal stability of the grown multi-layer, we followed a different 
approach compared to the straightforward heating of the multi-layer as 
described in Ref.~\citep{Krawietz1995}. There it is found that the reflectivity 
of the multi-layer is destroyed upon heating beyond 300$^{\circ}$C, explained 
by the formation of carbides which decompose upon heating at higher 
temperatures. 
%tussen
Here, we want to investigate the role of substrate temperature during 
deposition instead of post-processed as a thermal treatment is known to 
increase 
the thin film quality \citep{SarathKumar2013b, Tite2014}. We therefore use a 
growth temperature for the C layer of RT, 300$^{\circ}$C, 500$^{\circ}$C and 
700$^{\circ}$C. 
As film thicknesses of 1.2~nm (Ni) and 6.7~nm (C) are hard to measure by Cu 
K-$\alpha$ X-rays as the Kiessig fringes are very wide in 
2$\theta$, and one typically hits the noise floor 
of the detector for angles beyond 5$^{\circ}$, we study thicker C 
(75~nm) and Ni (20~nm) films here, in order to make it easier to determine the 
Kiessig fringes and to enable for more pronounced diffusivity through 
the multi-layer. 
Characterization by X-ray diffraction and reflectivity is plotted 
in Fig.~\ref{fig:2}(a) and (b).
%tussen
In Fig.~\ref{fig:2}(a) we plot the X-ray diffraction curve for a 
multi-layer grown at RT. The pattern reveals no specific 
crystallographic preferential orientation for Ni(111) at 44.6$^{\circ}$ 
\citep{Miller2012}, Ni(200) at 51.8$^{\circ}$ or Ni(220) at 76.4$^{\circ}$ 
in agreement with JCPDS No:04-0850.
Similar, the X-ray diffraction pattern shows no preferential 
crystallographic orientation for a crystalline C film, known to result in a 
weak and broad peak around 20$^{\circ}$ \citep{SarathKumar2013, 
Ruammaitree2013}. 
The peak at 33$^{\circ}$ and 69$^{\circ}$ 
corresponds to the silicon wafer, used as a substrate. In contrast to 
Ref.~\citep{Krawietz1995}, we do not observe any indication for 
crystalline Ni(111) nor Ni(200) formation in our curves, although the layers of 
the multi-layer described here are thicker and the effect is therefore expected 
to be 
more pronounced. 

The diffraction patterns, see 
Fig.~\ref{fig:2}(a), do not show any formation of carbides (Ni$_3$C) nor highly 
crystalline Ni structures as compared to Ref.~\citep{Krawietz1995}. From 
the bulk phase diagram \citep{Singleton1989}, these nickel carbide structures 
are anticipated to appear in a narrow temperature window \citep{Krawietz1995}, 
depending on the Ni/C ratio at the interface. As the temperature reading used 
in our experiments might be underestimated, we might have been beyond this 
temperature window, resulting in the lack of nickel carbide structures.

%Figure 2: XRD reveals no crystallinity, curve a) Ni curve b) RT grown C c) 
% % 300C grown C d) 500C grown e) 700C grown
% 160204A
% 160204B
% 160204C
% 
% 161028A/B/C/D 100nm C/ 20nm N

In Fig.~\ref{fig:2}(b) we plot the corresponding X-ray 
reflectivity curves, revealing clear Kiessig fringes, arising from the 
constructive interference between the X-rays  reflected from the film-vacuum, 
film-film and substrate-film interface. 
We find the reflectivity to be 
fully destroyed upon growth temperatures beyond 300$^{\circ}$C as can be seen 
by 
the absence of Kiessig fringes in Fig.~\ref{fig:2}(b), indicative of ill 
defined interfaces at the film-vacuum, film-film and substrate-film interface 
in agreement with literature \citep{Krawietz1995}. Obviously this is undesired 
for the application in mind.
Note that the critical angle decreases at increased growth 
temperature for the C layer, indicative of a decrease of the density of the 
multi-layer by about 25\%. Besides this, the periodicity of the Kiessig 
fringes is also shifting at 300$^{\circ}$C, suggesting C segregation into 
Ni described in literature. Partial segregation of the Ni would 
decrease the thickness of the reflective Ni layer, resulting in an increase in 
the width of the Kiessig fringes as observed here. This is hinting towards 
shifting of the interface between the C and Ni film, towards deeper depth.
To investigate the several interfaces in more detail and  study the cause for 
the destroyed reflectivity at growth temperatures beyond 300$^\circ$C, we make 
use of additional experimental techniques.

\begin{figure}
\includegraphics[width=8cm]{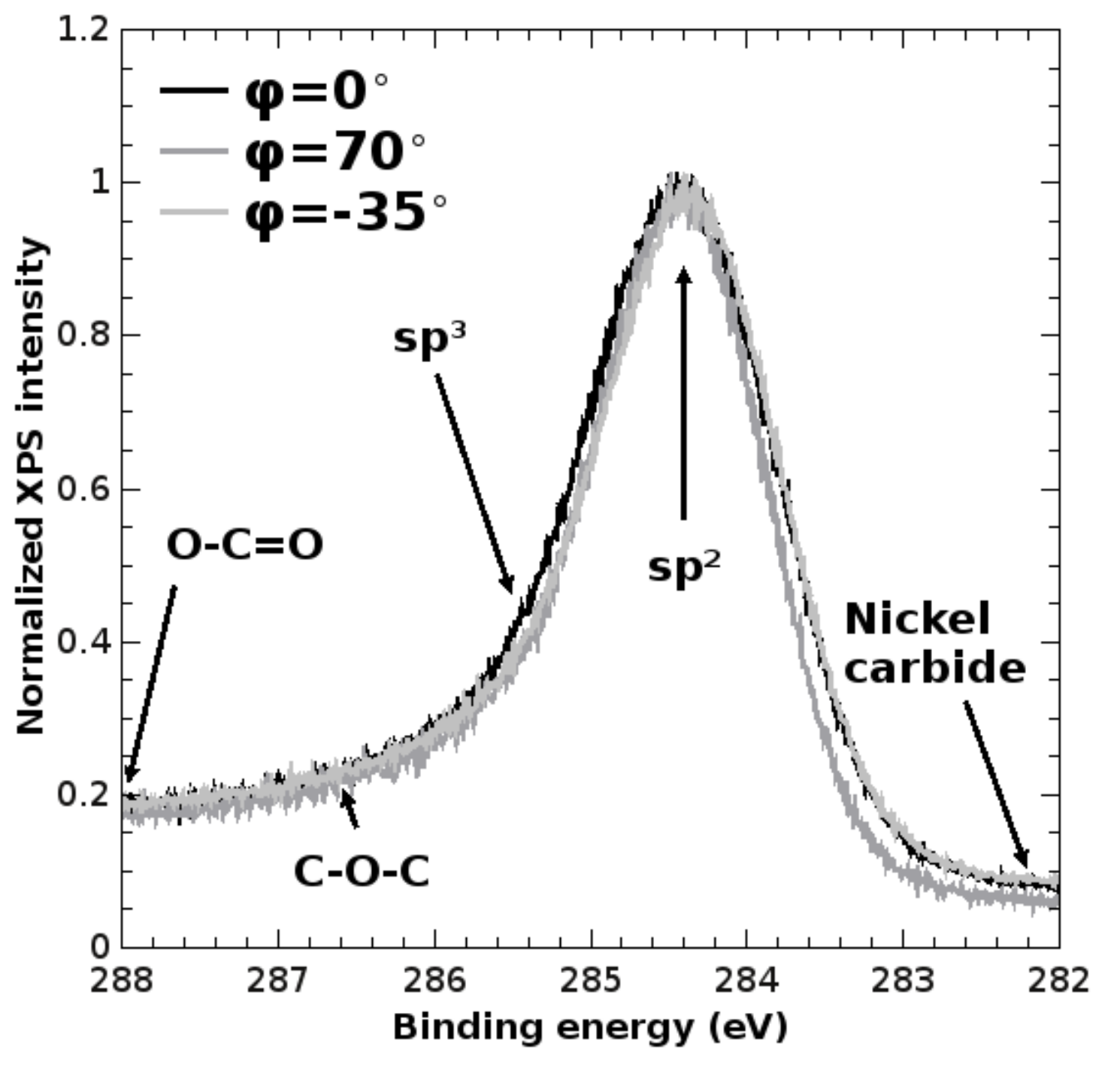}
\caption{X-ray Photoelectron Spectroscopy scan at several angles of incidence 
$\varphi$ around the C 1s peak for a 1.2~nm thick Ni, 6.7~nm thick C 
multi-layer grown at respectively RT and 700$^\circ$, revealing hardly any 
nickel carbide formation, corresponding to an energy of $\sim$282.8~eV.}
\label{fig:xps}
\end{figure}

To confirm the lack of nickel carbide formation at elevated temperature, we 
probed the surface chemical properties by depth-profiling X-ray photoelectron 
spectroscopy (XPS), see Fig.~\ref{fig:xps}. By variation of the angle of 
incidence, the estimated penetration depth is about 10~nm at $\varphi=0^\circ$ 
(for perpendicular incidence), 7~nm at $\varphi=-35^\circ$ and 3~nm at 
$\varphi=70^\circ$ incidence with respect to the substrates normal. The C 1s 
peak shows omnipresent contributions of the carbon 
$sp^2$ and $sp^3$ peaks at respectively 284.5~eV and 285.4~eV for all angles 
of incidence where the high ratio of $sp^2$:$sp^3$ is indicative for amorphous 
carbon \cite{Lascovich1991}. Nickel carbides, found around 282.8~eV 
\citep{Seo2017}, show only 
a faint signature when probing more surface sensitive. The C$-$O$-$C and 
O$-$C$=$O functional groups at respectively 286.5~eV and 288.7~eV seem absent 
as 
this film has been prepared and transferred to XPS in (ultra)high vacuum 
conditions. % These measurements support the observations of Ni segregation

\begin{figure}[h]
\includegraphics[width=8cm]{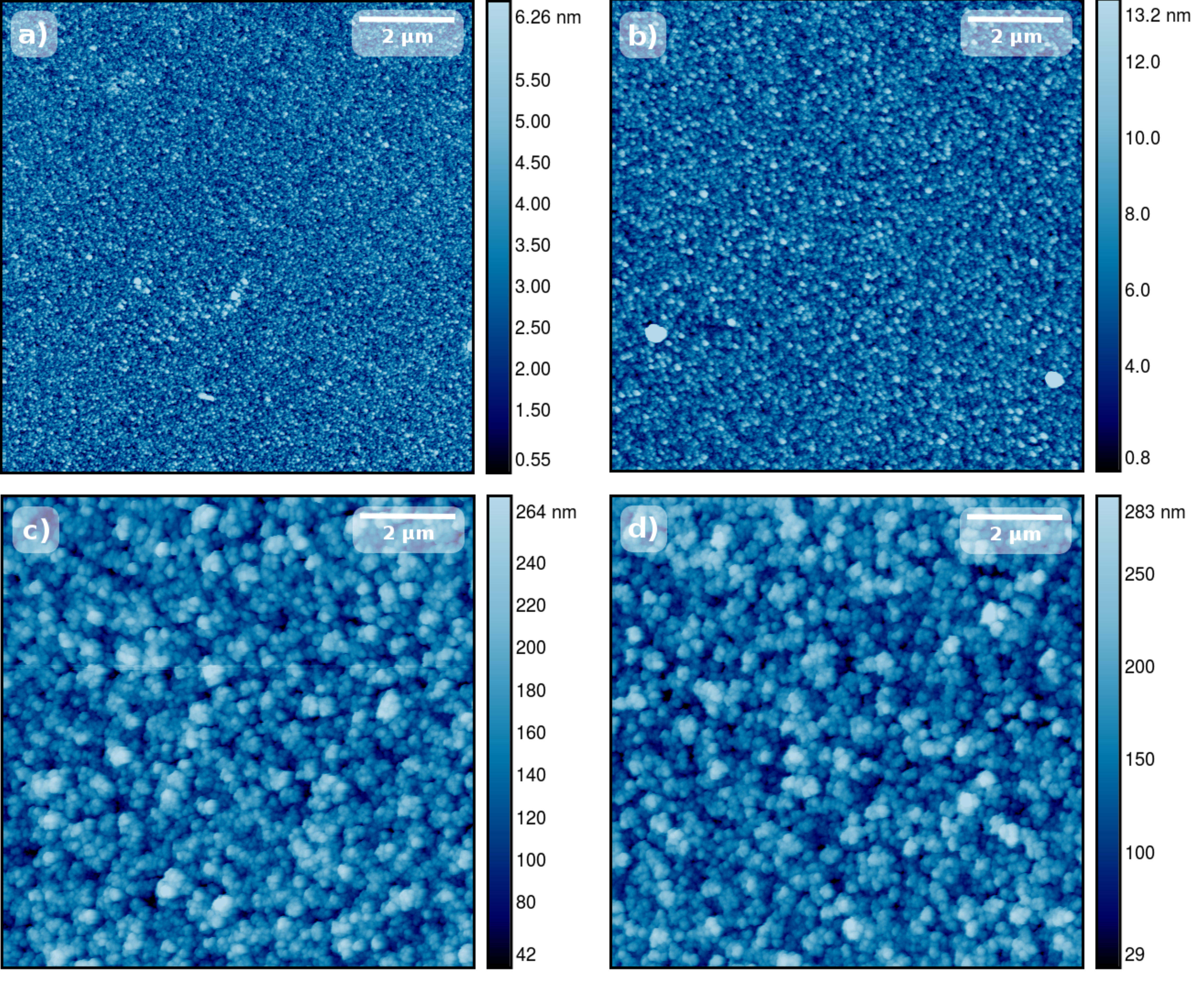}
\caption{(color online) TM-AFM images of 20~nm thick Ni film, grown at RT, 
followed by subsequent C film growth at RT (a), 300$^{\circ}$C (b), 
500$^{\circ}$C (c) and 700$^{\circ}$C (d).
}
\label{fig:5}
\end{figure}

As the loss of reflectivity can result from an ill defined film-vacuum, 
film-film and/or substrate-film interface, we investigate the film-vacuum 
surface morphology by AFM, see Fig.~\ref{fig:5}(a-d).
% and the buried film-film 
% and substrate-film interface by cross-section transmission electron microscopy 
% (XTEM), see Fig.~\ref{fig:4}. 
The measured surface 
roughness (RMS) by AFM for C films grown at RT is 0.89~nm, see 
Fig.~\ref{fig:5}. For 
increasing growth 
temperature, the roughness (RMS) increases to 2.1~nm at 300$^{\circ}$C, 37.6~nm 
at 500$^{\circ}$C and 45~nm at 700$^{\circ}$C. 
Krawietz et al. \citep{Krawietz1995} attribute the change in reflectivity 
beyond 
300$^{\circ}$C solely to the fragmentation of the Ni layers supported by the 
outdiffusion of C from the Ni layers.

%Figure 4: problem in reflectivity, afm fine, XTEM image

%Figure 5: XRR versus Temperature, next to AFM images

\begin{figure}
	\includegraphics[width=8cm]{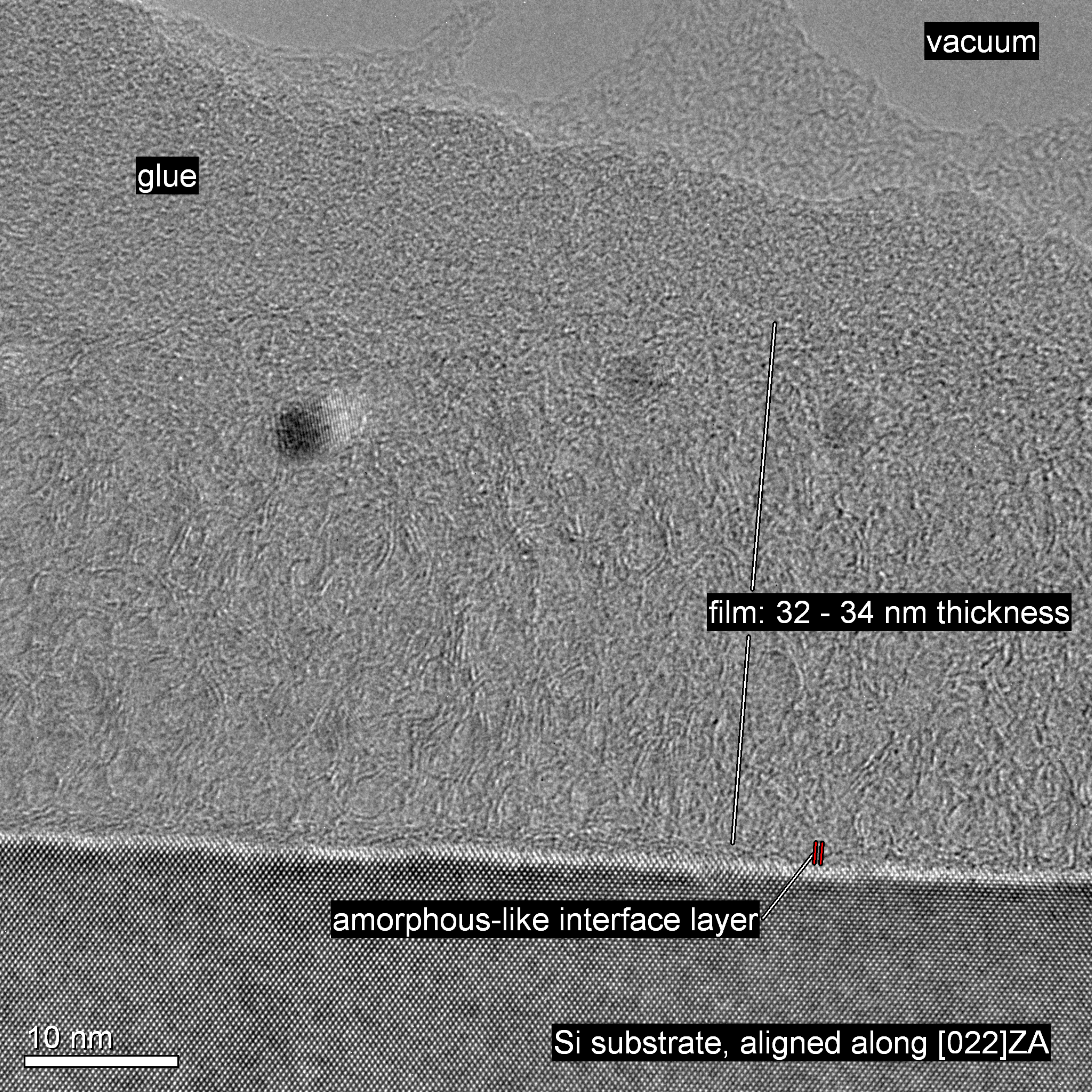}
	\caption{XTEM image of a multi-layer of 5 Ni/C stacks of 1.2~nm Ni and 
6.7~nm C. The highly crystalline substrate can be determined at the bottom, 
having an amorphous oxide layer on top. The following 32-34~nm contains the 
intermixed Ni/C layers, where intermixing occurred during growth at 
700$^{\circ}$C.}
	\label{fig:4}
\end{figure}

In order to investigate the different interfaces, we grow a multi-layer of 5 
Ni/C stacks containing a 1.2~nm thick Ni layer followed by a 6.7~nm thick C 
layer, where the Ni layer was grown at RT and the C layer was grown at 
700$^{\circ}$C. This sample was then perpendicular cut and grinded carefully to 
a thickness where XTEM could 
be performed. The resulting XTEM image is shown in Fig.~\ref{fig:4}. At the 
bottom of the image, the highly crystalline Si substrate is imaged along the 
[022]-axis. Above the Si substrate, a thin amorphous oxide layer can be 
discriminated. Above this, we find a thick layer of 32-34~nm in thickness, in 
good agreement with the deposited amount of Ni and C material. In this thick 
layer, we can not discriminate individual Ni and C layers. Instead, a layer 
of 
randomly oriented carbon nanotubes can be seen, with intermixed in it, 
crystalline clumps of Ni, similar to as described in Ref.~\citep{Peng2016}. 
The crystalline Ni clumps, having sizes up to 4-5~nm in diameter, suggest 
outdiffusion of C \citep{Krawietz1995}.

\begin{figure}
	\includegraphics[width=8cm]{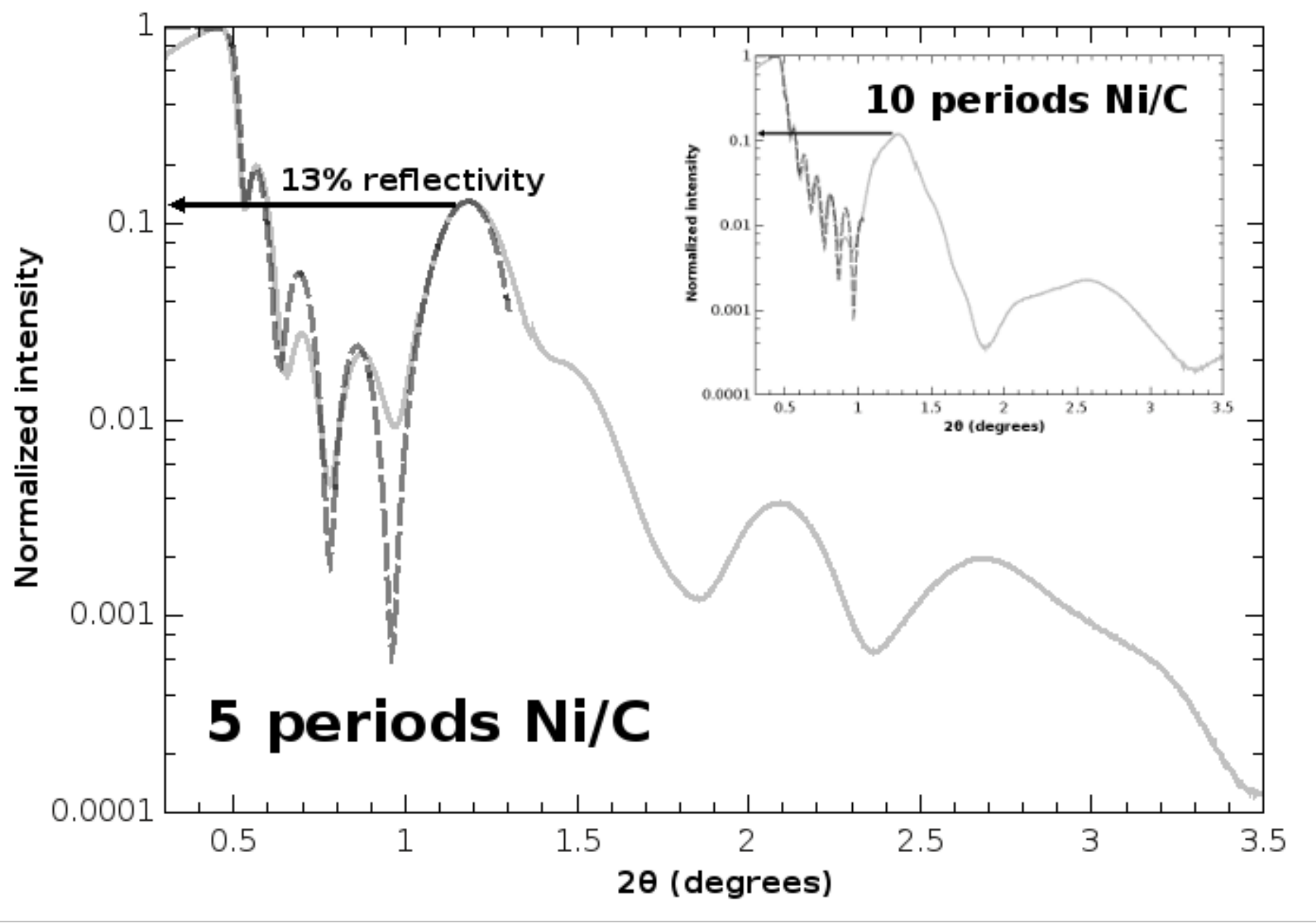}
	\caption{X-ray reflectivy curves for 5 pairs of Ni/C multi-layers  
and 10 pairs of Ni/C multi-layers (inset) with a 1.2~nm Ni and 6.7~nm C film 
thickness. The dashed curves in both graphs have been obtained by fitting as 
described in the text.}
	\label{fig:8}
\end{figure}

% added in Revision
To study the feasibility of creating high reflectivity multi-layers by PLD 
growth, we investigate the resulting reflectivity of several pairs of 
multi-layers grown.
As the growth of multi-layers at elevated temperatures results in intermixing 
of Ni and C destroying its reflective properties, we summarize this paper by 
growing a 5 and 10 pairs Ni/C multi-layer at RT, both with a Ni film thickness 
of 1.2~nm and a C film thickness of 6.7~nm at the same laser energy densities 
and process pressures described before. 
In Fig.~\ref{fig:8} we plot the X-ray reflectivity curve for 5 pairs 
of Ni/C multi-layers. For 10 pairs of Ni/C multi-layers, see inset of 
Fig.~\ref{fig:8}. 

From calculated X-ray reflectivity curves at 8.05~keV the resulting 
reflectivity at a grazing angle of $\theta$=0.6$^\circ$ is 13.9\% for 5 pairs 
increasing to 42.6\% for 10 pairs of multi-layers. An increase of the 
number of pairs of multi-layers results into an increased reflectivity at this 
grazing angle, see also Fig.~\ref{fig:0}(a). Analyzing the experimentally grown 
pairs of multi-layers in Fig.~\ref{fig:8}(a) and (b), we find a reflectivity for 
the 5 and 10 pairs of respectively 13.0\% and 11.8\%. Note that the reflectivity 
for 5 pairs of multi-layers is close to the theoretically predicted value, 
whereas for an increasing number of pairs the reflectivity does not increase.
In order to understand this, we quantify the roughness for the film-vacuum and 
substrate-film interface, by modeling the system as film pairs of uniform 
(electronic) density on top of a uniform (electronic) dense substrate. Fitting 
was done by using the fitting parameters film thickness ($d_{Ni}$ and 
$d_{C}$), assuming constant thicknesses for all layers, interface 
roughness between Ni ($R_{rms}^{Ni}$) and C films 
($R_{rms}^{C}$) and Ni-substrate interface roughness 
($R_{rms}^{substrate}$), a background resulting from scattering and a 
normalization factor \cite{GenX}. Note that we fit up to 
limited 2$\theta$ angle (see the dashed line in Fig.~\ref{fig:8}) to ensure the 
dynamical scattering theory is applicable and stay far from the kinetical 
scattering regime \cite{Jankowski2017,Vlieg2012}.
From this we find a steep roughness increase going from 5 pairs 
($R_{rms}^{C}$=0.27~nm and $R_{rms}^{Ni}$=0.41~nm) to 10 pairs of Ni/C 
multi-layers ($R_{rms}^{C}$=1.3~nm and $R_{rms}^{Ni}$=2.1~nm) grown by PLD at 
RT. Note that for 5 pairs, the fitted roughnesses agree well with the studied 
AFM images, see also Fig.~\ref{fig:1}. For 10 pairs of multi-layers, the 
roughness of samples grown at RT appears to be dictated by the droplets as 
discussed in Table~\ref{tab:1}.

%

%% conclusions
\section{Conclusions}
Here, we investigated the PLD grown low roughness Ni/C multi-layers by use of 
RHEED, XRD, XRR, TM-AFM, XPS and XTEM. Thin Ni and C layers with 
roughnesses below $\leq$0.2~nm can be created by PLD at RT. However, due to the 
fast heating and cooling occuring during melting of the targets surface, small 
droplets during deposition can not be avoided limiting the reflectivity at 
grazing angles for increased number of pairs of multi-layers.
Growth at temperatures of about 300$^{\circ}$C results in 
intermixing of Ni and C, destroying the reflective properties of the 
multi-layer. XTEM image analysis of multi-layers grown at elevated 
temperatures reveals (crystalline) Ni clumps intermixed with C.
For future applications in hard X-ray and soft $\gamma$-ray focusing 
telescopes, the stability of multi-layers under extreme (temperature) 
conditions 
is required and needs further investigation in order to find technological 
solutions.

% \acknowledgments
\section{Acknowledgments}
This work is part of the research programme of NanoNext NL project 7B 
Multilayered and artificial materials, project 12 and the research programme of 
the Foundation for Fundamental Research on Matter (FOM), which is part of the 
Netherlands Organisation for Scientific Research (NWO). 
Cross-section TEM imaging has been performed by Dr. Rico Keim. 
The author is gratefull to Prof.~Dr.~Ing.~G.~Rijnders for his permanent support.

\section{References}
\bibliographystyle{elsarticle-num}

% \bibliography{references.bib}
% \end{document}

\end{document}